\documentstyle[preprint,aps]{revtex}
\begin{document}

\def\a{\alpha}
\def\bs{\bigskip}
\def\cl{\centerline}
\def\ra{\rightarrow}
\def\al{\alpha}
\def\be{\beta}
\def\ga{\gamma}
\def\de{\delta}
\def\la{\lambda}
\def\bu{{\bf u}}
\def\bv{{\bf v}}
\def\bw{{\bf w}}

\def\cl{{\cal L}}
\def\si{\sigma}
\def\ur{\nearrow\ }
\def\ul{\nwarrow\ }
\def\dr{\searrow\ }
\def\dl{\swarrow\ }
\def\up{\uparrow\ }
\def\both{\leftrightarrow\ }

\def\bege{\begin{equation}}
\def\ende{\end{equation}}

\draft
\title{Lattice Statistics in Three Dimensions: Exact Solution of Layered Dimer
and Layered Domain Wall Models}
\author{V. Popkov\footnote{Permanent
address: Institute for Low Temperature Physics, Kharkov, Ukraine.}
and Doochul Kim}
\address{ Department of Physics and
Center for Theoretical Physics, Seoul  National University,\\
	 Seoul 151-742, Korea}
\author{H. Y. Huang and F. Y. Wu}
\address{Department of Physics,
		 Northeastern University, Boston, Massachusetts 02115}
\maketitle
\begin{abstract}
Exact analyses are given for two three-dimensional lattice
systems: A system of  close-packed
dimers placed in  layers of honeycomb lattices
and a  layered  triangular-lattice interacting domain wall model,
both  with  nontrivial interlayer   interactions,.
   We show that 
both models are equivalent 
to a 5-vertex model on the square lattice with 
interlayer vertex-vertex interactions. 
Using the method of Bethe ansatz,
a closed-form expression for the free energy is obtained and analyzed.
We deduce the exact phase diagram  and determine
 the nature of the phase transitions  as
a function of the strength of the interlayer
  interaction.
 
\end{abstract}
\bigskip

\section{Introduction}

An important milestone in the field of exact solutions of 
lattice-statistical systems is the solution of close-packed
dimers on planar lattices obtained by
Kasteleyn\cite{kas}
and by Fisher \cite{fisher}. 
  However, there has since been very little progress in extending
the dimer solution to  higher dimensions.
To be sure, Bhattacharjee et al.\cite{bhat}
have studied dimers on a certain 
three-dimensional (3$D$) lattice using numerical
means, and two of us\cite{wuhuang} have solved
a vertex model in arbitrary $d$ dimension, a solution
which also solves a dimer problem in $d$ dimension.
In the latter case, however, the dimer model involves 
unphysical negative statistical weights.

In a recent Letter \cite{5vl1},
hereafter referred to as I, three of us reported  on the
solution of a 3$D$ dimer system as an instance of a more 
general class of soluble 3$D$ lattice-statistical problem.
In contradistinction with other exactly solved 
3$D$ systems\cite{baxter,bb} which invariably involve negative
Boltzmann weights, the formulation reported in I,
which generalizes other  
special cases reported elsewhere  \cite{Popkov},
 marks  the success of solving   
a 3$D$ lattice-statistical model 
with strictly positive  Boltzmann weights.
In this paper we present  details of the analysis.
In addition, we 
show also  that our solution solves a layered domain wall model with 
interlayer interactions.

This paper is organized as follows. In Sec. II we define a layered dimer 
system with interlayer interactions and its equivalent layered 5-vertex model.  
The description of an equivalent layered domain 
wall model is given in Sec. III.
The free energy of the 3$D$ system is analyzed in Sec. IV with
the phase diagram obtained in Sec. V.  The critical behavior 
is deduced in Sec. VI.
Finally, in Sec. VII, we  discuss  the occurrence of  infinite
degeneracy  of orders in the system.
 
\section{A layered dimer system and the equivalent 5-vertex model}
 
Consider  a  3$D$ lattice  ${\cal  L}$ consisting of $K$ layers of honeycomb
lattices stacked together as shown in Fig. 1.  
Each layer of ${\cal  L}$ is an honeycomb  dimer lattice in which
dimers with weights $u, v, w$
are placed in the three respective  lattice directions.
 The dimers are close-packed within each layer
and  interact with an interlayer interaction shown in
Table I which gives, for example, 
the interaction energy $2h/3$,
and hence a Boltzmann factor $e^{-2h/3}$,
 between a $u$  dimer in the $k$th layer and
	 a $v$ dimer in the $(k+1)$th layer.
This completes the description of our 3$D$ dimer system.
Since a perusal of Table I shows that 
the negation of $h$ corresponds to the interchange of the layers
 $k$ and $k+1$, we can without loss of generality  take $h\geq 0$.

The honeycomb dimer system can be formulated as a 5-vertex model on a
square lattice \cite{wu68}.
This can be seen by 
 drawing the honeycomb lattice in the form of
a ``brick-wall'' as shown in Fig. 2.
The shrinking of each box containing two lattice points connected
by a $w$ edge into a point then converts 
the honeycomb lattice into a square lattice.
By regarding the presence of a $u$
or $v$  dimer on the remaining honeycomb edges as being 
    bonds,
each dimer configuration is then  mapped into
a vertex configuration of a  5-vertex model, and vice versa. 
The resulting five vertex configurations and weights\cite{wu68}
are shown in Fig. 3.
 
Furthermore, the interlayer dimer interaction 
 leads to an interlayer vertex interaction.
It turns out that the interlayer vertex-vertex interaction corresponding to 
Table 1 is not unique.  To deduce a useful interlayer vertex-vertex interaction
we first modify
Table 1 by replacing the
$uu$ and $vv$ entries  by $2\epsilon h$,
where $\epsilon = +1 \>(-1)$ for sites in sublattice A (B).  
Since two
     interacting $uu$ or $vv$ dimers are always parallel covering a pair of
	 A and B sites, this replacement
 does not alter the overall interaction energy.
A little algebra then shows that the dimer interaction
of the modified Table I leads to the interlayer vertex interactions shown in
Table II.
 Thus, we have at hand a layered 5-vertex model with
a particular interlayer vertex-vertex interaction.

Let each square lattice be of size $M\times N$, with
$M$ sites in a column
and  $ N$ sites in a row. This corresponds to 
$MNK$ dimers on ${\cal L}$.
Label sites of the layers  of square lattices by indices
$\{m,j,k\}$, with $m=1\dots M$, $j=1\dots N$ and $k=1\dots K$.  
Denote the vertex weight at site $\{m, j, k\}$ by
$W_{mjk}$, and denote the interaction 
Boltzmann weight between vertices
$\{m, j, k\}$ and $\{m, j, k+1\}$ 
as given in  Table II by $B_{mjk}$.
Then, it is our goal to evaluate the partition function
\begin{equation}
Z_{MNK} = \sum_{\rm config.}\prod_{k=1}
^K \prod_{m=1}^M \prod_{j=1}^N \biggl(B_{mjk}W_{mjk}\biggr) \label{part}
\end{equation}
where the summation is taken over all dimer, or vertex, configurations,
and the {\it per-dimer} free energy
\begin{equation}
f=K^{-1} \lim_{M,N\to\infty} (MN)^{-1}
\ln Z_{MNK}. \label{free}
\end{equation}
For simplicity, we shall assume $K=3\times$ integers.
We also assume periodic boundary conditions.

To write the interlayer vertex
interactions of Table II in the form of $B_{mjk}$,
 we introduce  variables 
  $\alpha_{mjk}= \pm 1$ and
$\beta_{mjk}=\pm 1$, respectively, for the horizontal and vertical edges
within the $k$th layer and originating from the site $\{m,j,k\}$ in
the direction of, say, decreasing $\{m,j\}$, such that $\alpha_{mjk}=+1\>
(-1)$ corresponds to the edge having a bond
 (empty).
It is then straightforward to verify that the vertex-vertex 
interactions in Table II
can  be  written as 
\begin{equation}
\varepsilon=-h \bigl( \alpha_{j} \tilde \beta_{j}-
\tilde \alpha_{j+1} \beta_{j}' \bigr) -{ h\over 3} \bigl( \alpha_j-\tilde 
\alpha_{j+1} \bigr) -{h\over 3} \bigl( \tilde \beta_j -\beta_j^\prime \bigr) ,
\label{eh}
\end{equation}
where
  we have, for convenience, suppressed the
subscripts $m$ and $k$ by adopting the notation 
\begin{equation}
\beta_{m+1,j,k} \rightarrow \beta_j', \hskip 1cm
\beta_{m,j,k+1}\rightarrow \tilde \beta_j, 
\label{convention}
\end{equation} 
and similarly for the $\alpha$'s.
Now the second and third terms in (\ref{eh}) are cancelled 
upon introducing this interaction into the overall partition function
(\ref{part}).  This   leads to  an effective 
 Boltzmann factor
\begin{equation} 
B_{mjk} = \exp \bigl(h \bigl( \alpha_{j} \tilde \beta_{j} -
          \tilde \alpha_{j+1} \beta_{j}' \bigr) 
		  \bigr) , 
\label{inter}
\end{equation}
which is to be used in (\ref{part}).

\section{A  layered domain wall model}
  
In this section we show that the layered dimer and 5-vertex models of 
the preceeding sections also describe a layered domain wall model with
interlayer interactions.

Consider a $3D$ lattice consisting of $K$ layers of triangular lattices
whose faces are elementary (up- and down-pointing) triangles.
Sites of the triangular lattices are occupied
by  Ising spins  $\sigma = \pm $ with the constraint that, around each
face of the lattice, there   are precisely   two spins
of the same sign and one spin of the opposite sign.
The allowed spin configurations are 
those of the ground state of an isotropic antiferromagnetic Ising model.
Furthermore, if one erases lattice edges connecting two 
spins of the same sign,  one arrives at a diamond (or rhombus) covering
of the triangular lattice.  This can be interpreted as 
 a dimer covering of the dual honeycomb lattice, by placing dimers 
connecting the two dual lattice points
on the elongated diagonal of each rhombus.
It is clear that the mapping between the spin configurations and
the diamond and dimer coverings  is      two to one.  Indeed,
 this mapping 
has been used   to extract  the 
solution of the honeycomb dimer lattice
from the  Ising ground state \cite{henk}. 
 
The spin 
  configurations  can also be viewed as representing 
domain wall configurations  \cite{henk,NohKim}.
   This mapping   is most conveniently seen   \cite{NohKim}
from the associated diamond covering scheme.
 If one attaches strips to those  diamonds  oriented in two
of the three possible directions as shown in Fig. 4, then the strips 
form continuous lines and propagate in a
zigzag but generally vertical direction,  which
 can be interpreted as
representing domain walls. 
(Cf. Figs. 2 and 4 of
\cite{NohKim} for a typical domain wall configuration.) A spin configuration
is thus mapped into  a domain wall
configuration.  Specifically, the triangular faces of the lattice
can be in one of the six ``strip" configurations shown in Fig. 5,
and the domain wall model is defined by associating weights to
the triangles as shown.

Next we introduce  interlayer domain wall interactions.
Shift the $(k+1)$-th layer by half lattice constant to the left with respect to 
the $k$-th layer so that  the up-pointing (down-pointing)
triangles in the layer $k$ 
 will be   adjacent to down-pointing (up-pointing) triangles 
in the layer $k+1$. 
Let two adjacent triangular faces
 in planes $k$ and $k+1$ interact with an energy   shown in
Table III.
 Together with the  triangle weights given in Fig. 5, this completely defines the 
layered domain wall problem.
 More precisely, the partition function for the domain wall problem is
now given by (\ref{part}),
where the summation extends over all domain wall
configurations,    with $W_{mjk}$ representing
the product of the triangle weights 
given in Fig. 5 and $B_{mjk}$ the interlayer interaction given 
by Table III.

The mapping of domain wall configurations to five-vertex arrow 
configurations has been 
given in \cite{NohKim}, where the triangular lattice 
was deformed into a square lattice by tilting it clockwise, 
leading to a 5-vertex model with $\omega_3=0$ (instead of 
$\omega_1=0$ as in Fig. 3).  For the present paper, we deformed
the triangular lattice  by tilting it counterclockwise. Then, the vertex 
weights reduce exactly to those given in Fig. 3.

To obtain an explicit form for $B_{mjk}$,
it is   straightforward to
verify that,
in the language of layered 5-vertex model, 
the interaction of Table III can be written as
 \begin{equation}
\varepsilon=-h \bigl( \alpha_{j} \tilde \beta_{j}'-
\tilde \alpha_{j+1} \beta_{j} \bigr) +{h} \bigl( \alpha_j-\tilde 
\alpha_{j+1} \bigr) -{h} \bigl( \tilde \beta_{j}' -\beta_j \bigr).
\label{penergy}
\end{equation}
  Again the second and third terms in (\ref{penergy}) are cancelled 
in the overall partition function
(\ref{part}),
so that the effective interaction Boltzmann factor now assumes the form
\bege
B_{mjk} ={\rm exp} \bigl( h \bigl( \alpha_{j} \tilde \beta_{j}'-
\tilde \alpha_{j+1} \beta_{j} \bigr) \bigr) ,
\label{b_domain}
\ende
which differs slightly from (\ref{inter}) for 
the dimer problem.  However, repeating precisely
the same line of argument as in I, one 
can show that the interlayer interaction (\ref{b_domain})
leads to precisely the same free energy (\ref{free2})
and (\ref{free1}) given below. 
Thus, the domain wall problem
(with interlayer interactions of Table III) is completely 
equivalent to  the dimer
system (with  interlayer interactions of Table I).

\section{The free energy}

In the preceeding sections we have established 
the complete equivalence of the  layered
dimer and domain wall problems, and their further equivalence with
a layered 5-vertex model.
 In this section we analyze the free energy of the layered 5-vertex problem.
For simplicity we use the language of the dimer system.

 It has been shown in I that
the layers of 5-vertex models with  interlayer interaction
(\ref{inter}) can be solved by applying
a transfer matrix in the vertical direction and  a global
Bethe ansatz consisting of the usual Bethe ansatz within each layer.
 This leads to the following expression for the  free energy
\begin{equation}
f(u,v,w,h) = \max_{-1\leq y_k\leq 1} f(\{y_k\}) , \label{free2}
\end{equation}
where
\begin{equation}
f(\{y_k\}) = \ln u + {1\over {K}} \sum_{k=1}^K
   {1\over {\pi }} \int_{0}^{ \pi(1-y_k)/2} \ln \biggl|
    {w\over u} + {v\over u} e^{2h(y_{k+1} -y_{k-1})} e^{i\theta} 
     \biggr| d\theta.
\label{free1}
\end{equation}
Here, 
$$
y_k=\frac{1}{N} \sum_{j=1}^N \beta_j
   =\frac{1}{N} \sum_{j=1}^N \beta_j^\prime ,
$$
is a quantity conserved from row to row (of vertical edges) in the $k$th layer
square lattice.
Specifically, we have $y_k=1-2n_k/N$,
where  $n_k$ is the number of vacant edges in a row.
Analysis  leading to (\ref{free1}) has been given in I and will not be
reproduced here.

 It is clear that for large
$u$, $v$ or $w$, the system is  frozen with complete
ordering of $u$, $v$, or $w$ dimers in all layers, and hence the free energies
\begin{eqnarray}
f_U &=& \ln u, \hskip 2cm  U \>\>{\rm phase} \nonumber \\
f_V  &=& \ln v, \hskip 2cm V \>\>{\rm phase} \nonumber  \\
f_W  &=& \ln w, \hskip 1.9cm W \>\>{\rm phase}. \label{fw}
\end{eqnarray}
 These are frozen orderings which we  refer
to as the $U$, $V$, and $W$ phases, respectively.
   For 
large $h$, it is readily seen from Table 1 that the 
energetically preferred state is the one in which each 
layer is occupied by one kind of dimers, $u$, $v$, or $w$,
and that the layers are ordered in the sequence of 
$\{u,w,v,u,w,v\cdots\}$. This ordered phase is referred to
as the $H$ phase with the free energy 
 \begin{equation} 
f_H = {1\over 3} \ln \bigl( uvw e^{4h} \bigr), \hskip 1.5cm
  H \>\>{\rm phase}
\end{equation}
obtained from a perusal of Table 1.

For any layer with $y_k=+1$ the corresponding integral
in (\ref{free1}) vanishes, and for $y_k=-1$ the integral can be evaluated
using the integration formula
\begin{equation}
{1\over \pi} \int_0^\pi \ln \bigl| A +B e^{i\theta} \bigr| d\theta
= {\rm max} \bigl\{ \ln |A|, \ln |B| \bigr\}. \label{integral}
\end{equation}
Therefore, the free energy ({\ref{free1}) can be explicitly evaluated if
 $y_k=\pm 1$ for all $k$.  Further discussion of this case
will be given in
Sec. VII.
 
    We have carried out
analytic as well numerical  analyses of the
free energy (\ref{free1})  for fixed $u,v,w,h$, and
it was found that the set  $\{y_k\}$ which gives the extremum
value  in (\ref{free2})  always repeats in multiples of 3, 
 namely, satisfying \cite{remark1}
$$
y_{k+3} = y_k. 
$$
The following extremum sets of $\{y_k\}$ are found:

\medskip
\noindent
1.  $\{y_1, y_2, y_3\} = \{1,1,1\}$:
In this case we have all $y_k=1$, and hence from (\ref{free1})
\begin{equation}
f=f_U.  \hskip 2cm U \>\>{\rm phase}
\end{equation}
 This gives rise to the $U$ phase.

\medskip
\noindent
2. $\{y_1, y_2, y_3\} = \{-1, -1, -1\}$:
In this case we have  all $y_k=-1$, and hence from ({\ref{free1})
\begin{eqnarray}
f &=& \ln u + {1\over \pi} \int_0^\pi \ln \biggl| {w\over u} +{v\over u}
e^{i\theta}
\biggr| d \theta \nonumber \\
&=& f_W, \hskip 2cm w>v \hskip 1cm W \>\>{\rm phase} \nonumber \\
&=& f_V, \hskip 2cm v>w.\hskip 0.95cm V\>\>{\rm phase}
\end{eqnarray}
 This gives rise to the $W$ and $V$ phases.

\medskip
\noindent
3. $\{y_1, y_2, y_3\} = \{1, -1, -1\}$:
Substituting this sequence of $y_k$ values into ({\ref{free1}) and 
making use of (\ref{integral}) in the resulting expression, one obtains
\begin{eqnarray}
f&=& \ln u  + {1\over {6\pi}} \int_{-\pi}^\pi \ln \biggl| {w\over u} +
              {v\over u} e^{2h} e^{i\theta}\biggr| d \theta
             + {1\over {6\pi}} \int_{-\pi}^\pi \ln \biggl| {w\over u} +
              {v\over u} e^{-2h} e^{i\theta} \biggr| d \theta \nonumber \\
  &=& {1\over 3} f_U + {2\over 3} f_W , \hskip 2.5cm 
ve^{-4h}< ve^{4h}<w \label{fh1} \\
  &=& {1\over 3} f_U + {2\over 3} f_V , \hskip 2.5cm w< ve^{-4h}<ve^{4h} \label{fh2} \\
   &=& f_H .  \hskip 3.5cm   ve^{-4h} < w <  ve^{4h}, \hskip 0.5cm H 
\>\> {\rm phase} \label{fh3}
\end{eqnarray}
Now the free energies (\ref{fh1}) and (\ref{fh2}) 
can be discarded since they are always smaller than
the largest of $\{f_U, f_V, f_W\}$.
Thus, this set of
$\{y_k\}$ leads to 
a frozen ordering for  sufficiently large $h$ as indicated in
(\ref{fh3}), which is the $H$ phase.

 \medskip
\noindent
4. $\{y_1, y_2, y_3\} =\{y, y, y\}$:
In this case all $y_k = y$, where $y$  maximizes the  free
energy  (\ref{free1}).  
 Then,  substituting  $y_k=y$ into (\ref{free1}) and 
carrying out the maximization in (\ref{free2}) by a straightforward
differentiation with respect to  $y$, one obtains 
\begin{equation}
f = f_Y(y_0) \equiv  \ln u + \frac{1}{\pi}\int_0^{\pi(1-y_0)/2}
   \ln \biggl|{w\over u} +{v\over u} e^{i\theta}\biggr| d\theta ,
   \hskip 1cm
Y\>\>{\rm phase} \label{fy}
\end{equation}
where the extremum $y_0$ is given by
 \begin{equation}
{\pi\over 2} (1-y_0) =\cos^{-1} \biggl[{{u^2-w^2-v^2}\over {2wv}}
\biggr]. \label{y0}
\end{equation}
This is a disorder phase which we
 refer to  as  the $Y$ phase.
Despite its apparent asymmetric appearance, the 
free energy $f_Y(y)$ is actually
symmetric in  $u, v, w$.  
Note that, for large $v\sim w$, we have 
\begin{equation}
{\pi\over 2} (1-y_0) = \pi - \theta_0, \label{s0}
\end{equation}
where $\theta_0$ is small and given by
\begin{equation}
\theta_0 ^2 = [u^2 -(w-v)^2]/wv.  \label{t0}
\end{equation}

\medskip
\noindent
5. $\{y_1,y_2,y_3\} = \{ y_1, -1,-1\}$:
This is the $H$ phase with the  $u$ layers replaced by 
  layers with $y_k=y_1$,  so that 
the layer ordering is $\{ y_1,w,v, y_1,w,v \cdots\}$.  
This is a partially ordered phase which
we refer to    as the $I_u$ phase.
Again, the substitution of these values of $\{y_k\}$ into (\ref{free1}) and
a straightforward maximization yield, after using (\ref{integral}),
 \begin{eqnarray}
f=f_{I_u}({ y_{10}})&=& {1\over 3} \biggl[2f_W +  f_Y(y_{10})\biggr],
 \hskip 1cm w>ve^{4h(1+y_{10})} 
    \label{freeu1} \\
&=&  {1\over 3}\biggl[ 2f_V +  f_Y(y_{10})\biggr], \hskip 1cm
w<ve^{-2h(1+y_{10})} 
    \label{freeu2} \\
&=&
\frac{1}{3} \biggl[\ln \bigl(vwe^{2h(1+{ y_{10}})}\bigr)
        + f_Y(y_{10}) \biggr] , \nonumber \\
& & \hskip 1.5cm ve^{-2h(1+y_{10})}<w< ve^{2h(1+y_{10})},
\hskip 0.4cm I_u \>\>{\rm phase} \label{freeu}
\end{eqnarray}
 where 
$f_Y(y_{10})$ is defined in (\ref{fy}), with the
extremum $y_{10}$ given by
 \begin{equation}
{\pi\over 2} (1-y_{10}) =\cos^{-1} \biggl[{{u^2e^{8h}-w^2-v^2}\over {2wv}}
\biggr]. \label{y1}
\end{equation}
  The free energies (\ref{freeu1}) and 
(\ref{freeu2}) are discarded since 
they are always less than the largest of $\{f_W, f_V, f_Y(y_{10})\}$, and 
we have $f_Y(y_{10})<f_Y(y_0)$ by definition.
Therefore, the free energy of the $I_u$ phase is given by (\ref{freeu}).
Note that, for large $v\sim w$, we have 
\begin{equation}
{\pi\over 2} (1-y_{10}) = \pi - \theta_1, \label{s1}
\end{equation}
where $\theta_1$ is small and given by
\begin{equation}
\theta_1 ^2 = [u^2e^{8h} -(w-v)^2]/wv.  \label{t1}
\end{equation}

\medskip
\noindent
5. $\{y_1,y_2,y_3\} = \{1, y_2, -1\}$:
This is the $H$ phase with the  $w$ layers replaced by 
  layers with $y_{k+1}=y_2$,  so that 
the layer ordering is $\{ u,y_2,v,u, y_2,v \cdots\}$.  
We refer to  this  as the $I_v$ phase.
Due to the intrinsic symmetry of the interlayer interaction, the free energy
of the $I_w$ phase is the same as that of $I_u$, given in (\ref{freeu}), with
the cyclic permutation of $u\to w\to v \to u$.  Alternately, one can substitute
these $\{y_k\}$ values into (\ref{free1}) and
carry out the  maximization.  It can be verified that this leads to
 \begin{equation}
f=f_{I_w}({ y}_{20})=\frac{1}{3}\biggl[ 2 f_U +  f_V 
 +  {1\over{ \pi}} \int_{0}^{\pi(1-y_{20})/2}
 \ln \biggl|{v\over u} + {w\over u} e^{4h} e^{i\theta} \biggr|d\theta
\biggr],\hskip 0.5cm I_w\>\>{\rm phase} \label{freew}
\end{equation}
 with 
 \begin{equation}
{\pi\over 2} (1-y_{20}) =\cos^{-1} \biggl[{{u^2-w^2e^{8h}-v^2}\over
{2wve^{4h}}}
\biggr]. \label{y2}
\end{equation}

\medskip
\noindent
6. $\{y_1,y_2,y_3\} = \{1,-1, y_3\}$:
This is the $H$ phase with the  $v$ layers replaced by 
  layers with $y_{k+2}=y_3$,  so that 
the layer ordering is $\{ u,w,y_3,u,w, y_3  \cdots\}$.  
We refer to  this  as the $I_v$ phase.
Again, the free energy of the $I_v$ phase can be written down by  symmetry.
Alternately,
the substitution of these values of $\{y_k\}$ into (\ref{free1}) and
(\ref{free2})
 yields
 \begin{equation}
f=f_{I_v}({ y}_{30})=\frac {1}{3} \biggl[2f_U +  f_W
  +  {1\over{ \pi}} \int_{0}^{\pi(1-y_{30})/2}
 \ln \biggl|{v\over u} e^{4h} + {w\over u} e^{i\theta} \biggr|d\theta
\biggr], \hskip 0.5cm I_v\>\>{\rm phase}\label{freev}
\end{equation}
 with 
 \begin{equation}
{\pi\over 2} (1-y_{30}) =\cos^{-1} \biggl[{{u^2-w^2-v^2e^{8h}}\over
{2wve^{4h}}}
\biggr]. \label{y3}
\end{equation}

\section{The phase diagram}

 Since the phase diagram must reflect the
$\{u,v,w\}$ symmetry of the
 interlayer interaction
given in Table 1,
 it is  convenient to introduce 
coordinates
\begin{equation}
X=\ln(v/w) \hskip1cm Y= (\sqrt{3})^{-1} \ln (vw/u^2)
\end{equation}
such that any interchange of the three variables $u$, $v$, and $w$
corresponds to a $120^\circ$ rotation in the $\{X, Y\}$ plane.
The phase boundaries are then determined by equating the free energies
of adjacent phases. The results are collected in Fig. 6.

The phase diagrams depend on the value of $h$ and are different in different
regimes.

$\bullet$ $h<h_0$:
For small $h$ the phase diagram is the same as  that of the $h=0$
noninteracting
2$D$ system, namely, the diagram shown in Fig. 6a. 
The phase boundary between the $\{U,V,W\}$ phases and the $Y$ phase,
which stays the same in all regimes below,
is  obtained by setting $y_0 = \pm 1$ in (\ref{y0})
where $f_U, f_V$ or $f_W$ is equal to $f_Y$.
These boundaries are 
\begin{equation}
u=|v\pm w|.  \label{yuvwboundary}
\end{equation}
In terms of the coordinates $X$ and $Y$,  
$u=v+w$ and $u=|v-w|$ read, respectively,
$$
 Y = {{-2}\over {\sqrt 3}}\ln[2\cosh(X/2)], 
\hskip 0.5cm  \>\> Y = {{-2}\over {\sqrt 3}}\ln[2\sinh(|X|/2)].
$$

$\bullet$ $h_0<h<h_1$:
As $h$ increases from zero, our numerical analyses 
indicate that the $H$ phase appears when $h$
reaches a certain value $h_0$.  The resulting phase diagram is shown
in Fig. 6b.    The phase boundary
between the $H$ and $Y$ phases is  given by
$f_H =f_Y$, or, explicitly
\begin{equation}
{1\over 3} \ln \bigl( uvw e^{4h} \bigr) = f_Y(y_0).\label{hyboundary}
\end{equation}
Thus, $h_0$ is obtained from ({\ref{hyboundary}) by 
 setting $u=v=w$ ($X=Y=0$) where the $H$ phase first appears.
This  yields $\pi(1-y_0)/2 = 2\pi/3$ and
 \begin{equation}
h_0 = \frac{3}{8\pi} \int_0^{2\pi/3}
               \ln ( 2+2\cos \theta ) d\theta =0.2422995 \cdots.
\label{fhfy1}
\end{equation}

$\bullet$ $h_1<h<h_2$:
As $h$ increases from $h_0$, the $I_u, I_v, I_w$ phases appears 
when $h$ reaches a certain value
$h_1$. The resulting phase diagram is shown in Fig. 6c.
Now the $I_u$ phase is the $H$ phase with the $u$ layers (with $y_k=1$)
replaced by layers with  $y_k=y_{01}$, the  boundary between
the two regimes is therefore given by
$y_{10}=1$ or, explicitly using (\ref{y1}),
\begin{equation}
w+v= ue^{4h}. \label{iuhboundary}
\end{equation}
The  boundary between the $I_u$ and the
$Y$ phases is
$ f_Y(y_0)=f_{I_u}(y_{10})$ or, explicitly,
\begin{eqnarray}
&&\ln u +{1\over \pi} \int_0^{\pi(1-y_0)/2} \ln \biggl|{v\over u} 
+{w\over u} e^{i\theta} \biggr| d\theta  \nonumber \\
&& \hskip 0.5cm
= {1\over 3} \biggl[ \ln \biggl(vwe^{2h(1+y_{10})}\biggr) + \ln u +
{1\over \pi} \int_0^{\pi(1-y_{10})/2} \ln \biggl|{v\over u} 
+{w\over u}e^{i\theta} \biggr| d\theta \biggr].
 \label{iuyboundary}
\end{eqnarray}
The boundaries of the $I_v$ and $I_w$ regimes can be written down similarly.

To compute the numerical value of $h_1$,
we note that the $I_u$ phase first appears at $v=w$ ($X=0$) when all three
phases $H$, $Y$ and $I_u$ coincide.   Therefore, $h_1$ is obtained by solving
(\ref{hyboundary}) and (\ref{iuhboundary}) for $v/u$ and $h$ at
$w=v$.  This leads to the value
$$
h_1 = {1\over 4} \ln \biggl({{2v}\over u} \biggr)= 0.2552479\cdots,
$$
where $v/u$ is the   solution of the equation
\begin{equation}
(1+y_0) \ln {v\over u} + {1\over 6} (1+3y_0)\ln 2
= {1\over \pi} \int_0^{\pi(1-y_0)/2} \ln(1+\cos\theta)d\theta
\end{equation}
with 
\begin{equation}
{\pi\over 2} (1-y_0) = \cos ^{-1} \biggl({{u^2}\over {2v^2}} -1\biggr).
\end{equation}

$\bullet$ $h_2<h<h_3$: 
As $h$ increases from $h_1$, it was found that the regimes $I_u$, $I_v$, and
$I_w$
extends to infinite when $h$ exceeds a certain value $h_2$.  The phase
diagram is shown in Fig. 6d. 
The value of $h_2$ can be deduced from (\ref{iuyboundary}) in its large
$w=v$ expansion. 
Setting $w=v$ in (\ref{iuyboundary}), introducing
(\ref{s0}) and (\ref{s1}) for large $w,v$,  and equating the coefficients
of $\ln(v/u)$ on both sides of the equation,
one obtains
\begin{equation}
{1\over \pi} (\pi -\theta_0) = {1\over 3} \biggl[2 +{1\over \pi} (\pi -
\theta_1)
\biggr],
\end{equation}
or, simply, $\theta_0=\theta_1/3$.  Now from 
(\ref{t0}), (\ref{t1}), and $w=v$, we have the expressions $\theta_0= u/v$ and 
$\theta_1=(u/v)e^{4h}$.  It follows that 
we have
$$
h_2 = (\ln 3) / 4 = 0.3816955\cdots.
$$
 
$\bullet$ $h_3<h<h_4$: 
As $h$ increases from $h_2$, it was found that the boundary of the
$H$ phase bulges toward the $U,V,W$ phases along the $30^\circ, 
150^\circ, 270^\circ$ lines, touching the $U,V,W$ boundaries
in these directions when $h$ equals a certain value
$h_3$.  
For $h>h_3$, the $H$ phase borders directly with the 
$U,V,W$ phases with respective boundaries
\begin{equation}
u^2 = vwe^{4h}, \hskip 1cm v^2 = wue^{4h}, \hskip 1cm 
w^2 = uve^{4h}. \label{huvwboundary}
\end{equation}
The size of these borders
grows while the $Y$ phase shrinks as $h$ increases.
The phase diagram in this regime is shown in Fig. 6e.
To determine $h_3$, we 
let the $\{H, Y\}$ phase boundary (\ref{hyboundary}) to touch
the $\{Y,U\}$ phase boundary
  $ u=w+v$ at $w=v$
(the $270^\circ$
direction).  Using (\ref{y0})  we have $y_0=1$, and it follows that 
(\ref{hyboundary}) becomes
$$
 {1\over 3} \biggl[ \ln(2v^3 e^{4h})\biggr] =  \ln (2v) +0 
$$
from which we find \cite{remark}
$$
h_3 = (\ln 2)/2 = 0.3465735\cdots.
$$

$\bullet$ $h>h_4$: 
As $h$ increases further from $h_3$, it was found that the 
 $Y$ phase disappears completely when $h$ exceeds a 
certain value $h_4$.  The
 phase
diagram in this regime is shown in Fig. 6f. 
To determine the numerical value of $h_4$, we note that the $Y$ phase
disappears
when the boundary $v=w+u$ between the $V$ and $Y$ phases coincides with the
boundary
(\ref{iuyboundary}) between the $I_u$ and $Y$ phases  at
large $w,v$.  Therefore, we again   expand (\ref{iuyboundary}) for large
$v,w$ but now subject to $v-w=u$.  Introduce (\ref{s0}) and (\ref{s1})
for the integration limits.
But now from (\ref{t0}) and (\ref{t1}) we have 
$\theta_0 = 0, \theta_1 = \gamma u/\sqrt{wv},$ where
\begin{equation}
\gamma =\sqrt {e^{8h}-1}. \nonumber
\end{equation}

Substituting (\ref{s0}) and (\ref{s1}) into (\ref{iuyboundary}) and 
making use of (\ref{integral}) and the relation (for $v>w$)
$$
{1\over \pi} \int_0^{\pi - \theta_1} 
 \ln \biggl| {v\over u} +{w\over u} e^{i\theta}\biggr|
d\theta
= \ln {v\over u}  - {1\over \pi}
\int_0^{\theta_1} 
 \ln \biggl| {v\over u} -{w\over u} e^{i\theta}\biggr|
d\theta,  
$$
one obtains   
\begin{equation}
\ln v  =
{1\over 3}\biggl[ \ln (vwe^{4h\theta_1/\pi}) + \ln v
  - {1\over {2\pi}} 
\int_0^{\theta_1} \ln \biggl({{v^2}\over {u^2}}+ {{w^2}\over {u^2}}
-{{2vw}\over {u^2}}\cos \theta \biggr)d\theta \biggr].\label{dh3}
\end{equation}
Since  $\theta_1$ is small, we can write $\cos \theta = 1-\theta^2/2$
in  the integrand, and the integral can be simplified by
introducing the change of variable  $y= \sqrt {wv}\theta/u$.
Introduce $w=v-u$  and expand  (\ref{dh3}) for large $w,v$ using,
for example, $\ln(vw) = 2\ln v - u/v$,
 the leading terms   of the order of
$\ln v$  are cancelled.  The next  terms
 including the integral  are of the order of
$O(u/v)$.
Setting the coefficient of these  term equal to zero, one obtains
\begin{eqnarray}
 4h\gamma &=& \pi + {1\over 2}\int_0^\gamma \ln (1+y^2) dy \nonumber\\
&=& \pi +{1\over 2} \gamma \ln(1+\gamma^2) - \gamma +  \tan^{-1} \gamma,
\nonumber
\end{eqnarray}
or, after using $1+\gamma^2 = e^{8h}$, 
\begin{equation}
\gamma -\tan^{-1} \gamma = \pi \label{gamma}
\end{equation}
whose solution gives $h_4$.  Specifically, we find

$$ 
h_4= {1\over 8 } \ln (1+\gamma^2) = 0.3816955\cdots.
$$ 

\noindent
{\it Phase diagram for the domain wall model}.

Since the domain wall model with weights given in Fig. 5 and interlayer
interactions of Table III is completely equivalent to the layered dimer
system, the phase diagram of the domain wall model is the same as that 
of the dimer model.
For example, the phase $U$ corresponds to a phase with no domain
walls, and that  the phase $H$ corresponds to 
a sequence of triplets of layers with no domain walls, maximal
density of domain walls consisting of elementary weights $\sqrt w$,
and maximal density of domain walls of  weights $\sqrt v$
(Cf. Fig. 5).

\section{The critical behavior}

In this section we 
determine the critical behavior  near all phase
boundaries.
 Since the free energies are  given by different 
analytic expressions in different 
phase regimes, one generally expects the first 
derivatives of the free energy (with respect to a temperature $T$, say)
be discontinuous.  This then leads to first-order
transitions.   However, if the first derivatives
of the  free energies happen to vanish on both sides of the boundary,
 then one has a continuous transition.
Applying this reasoning, we find
 all transitions   to be of first order,
except those
between the $\{U,V,W\}$ and $Y$ phases, and between the $\{I_u, I_v, I_w\}$
 and $H$
phases, which are found to be the 
same as the transition in the 5-vertex model \cite{wu68,wu67},
namely, a
continuous transition with a square-root singularity in
the specific heat.  This transition,
first reported by one of us in 1967 \cite{wu67},
is now known as the Pokrovsky-Talapov type
phase transition \cite{pokro}.
  
Regarding
$u,v,w$ and $e^h$ as Boltzmann factors, 
the ordered $U, V, W$ and $H$ phases  (with $y_k=\pm 1$)  have constant
free energies and hence zero
 first derivatives.
   Therefore,
we focus on the boundaries of these frozen regimes.

We have seen that
 the transition  between the $U,V,W$ phases and
the $Y$ phase  is the same as that occurring in the 2$D$ system,
which is  known \cite{wu68} to be continuous.
This fact can also 
be seen by expanding  the free energy   near $y_0$ as
\begin{equation}
f_Y(y) = f_Y(y_0) +(y-y_0)f_Y'(y_0) +{1\over {2!}} (y-y_0)^2f_Y''(y_0)
   +{1\over {3!}} (y-y_0)^3f_Y'''(y_0) +\cdots. \label{taylor}
\end{equation}
Using the expression of $f_Y(y)$ defined by (\ref{fy}), one sees that, indeed,
the first derivative $f_Y(y_0)$, $y_0=\pm 1$,  vanishes identically on the boundary
(\ref{yuvwboundary}) which is precisely $f_Y'(y_0) =0$.
  Furthermore, it is also seen that 
$f_Y''(y_0) \sim \sin [\pi(1-y_0)/2] =0$.  Therefore, the extremum of
$f_Y(y)$  (\ref{taylor}) occurs at $y=y_{\rm extrm}$ given by
\begin{equation}
y_{\rm extrm} -  y_0 = \pm
\sqrt{\frac  {2 f_Y'(y_0)}    {-f_Y'''(y_0)}   }
          \sim t^{1/2},
\end{equation}
where $t= |T-T_c|$,  $T_c$ being the critical temperature.
 Substituting this $y_{\rm extrm}$ into (\ref{taylor}), one obtains
\begin{eqnarray}
f_Y(y_{\rm extrm})&=&f_Y(y_0) \pm \frac{2}{3} f_Y'(y_0)
            \sqrt{\frac{2 f_Y'(y_0)}{-f_Y'''(y_0)}}
           \nonumber \\
           &=&f_Y(y_0) + c(u,v,w,h) t^{3/2}.
\end{eqnarray}
This leads to a square-root singularity in the specific heat,
which is a characteristic of the Pokrovsky-Talapov phase transition.
The key element leading to this result is 
the fact that the boundary of the frozen phases
is given
precisely by $f_Y'(y_0)=0$,
rendering  the first derivative of the free energy
to vanish at the boundary.
 
Applying the same analysis to 
the  $H$ and $I_u$ phases,  the boundary
is again given by
$ f_{I_u}'(y_{10}) =0$.
   Furthermore, it is also seen
that  $f_{I_u}''(y_{10}) =0$ identically.  It follows that the analysis
can be carried through exactly as given in the above, and one concludes
\begin{equation}
f_{I_u}(y_{\rm extrm})
            =f(y_{10}) + c_1(u,v,w,h) t^{3/2}. \nonumber
\end{equation}
This gives rise again  to a square-root singularity in the specific heat.
The consideration of
 the $I_v, I_w$ and the $H$ boundaries
can be carried out in a similar fashion.

\section{Degeneracy  of ordered states}

We discuss in this section the degeneracy of  ordered states.
  Particularly, we show that  
the system has a nonzero per-layer entropy
 on
 the boundaries between $H$ and $U,V,W$  phases.
 
We first establish the occurrence of a degeneracy from an energy consideration.
For this purpose it is sufficient 
to show that this is the case along the $\{H, V\}$ boundary
(\ref{huvwboundary}), namely,
\begin{equation}
e^{4h} = v^2/wu. \label{uvwh}
\end{equation}
We already know that, along 
the boundary (\ref{uvwh}), the following  layer orderings of the $H$ and
$V$ phase are degenerate,
 \begin{eqnarray}
&& \hskip 1cm ...222222222... \nonumber \\
&& \hskip 1cm ...(132)(132)(132).... \label{ord}
\end{eqnarray}
Here, for convenience, we have used the notations $\{1,2,3\}$ for $\{u,v,w\}$.
Generally, when $y_k=\pm 1$,  each layer contains
dimers of only one kind, $u,v$ or $w$.  Let  $\a_i$, $i=1,2,3$
denote the numbers  of  $u,v$ and $w$ layers, respectively, 
as a fraction of $K$, and $\a_{12}$
the fraction of adjacent pair of $u,v$ layers, etc.
Then, perusal of Table I shows that this leads to 
the per-site free energy
\begin{equation}
f= {1\over 2} \biggl(\a_1\ln u + \a_2\ln v +\a_3 \ln w \biggr) +{{2h}\over 3}
\biggl(
 \a_{13}+ \a_{32}+ \a_{21}- \a_{12}- \a_{23}- \a_{31}\biggr).\label{free5}
\end{equation}
Here, the $\a$'s satisfy the conservation rules $\a_1+\a_2+\a_3=1$,
$\sum_j \a_{ij} = \a_i$.   

Consider the following  ordering,
\begin{equation}
 ...2(1)3222(1)322222(1)32..., \label{o2}
\end{equation}
characterized by single $u$ layers separating strings of layers of the type
$wvvvvv...$ where there is at least one $v$ layer in each string.
 It is readily seen using (\ref{o2}) that for this ordering we have
\begin{eqnarray}
&& \a_{12} =\a_{23} =\a_{31} = 0\nonumber \\
&& \a_{21} =\a_{32} =\a_{13} =\a_1=\a_3 \equiv \a, 
\hskip 1cm \a_2 = 1-2\a. \label{o3}
\end{eqnarray}
It is then  a simple matter to substitute (\ref{uvwh}), 
(\ref{o3}) into (\ref{free5}), obtaining
$f=f_V$.  Thus, any layer ordering having the structure of (\ref{o2})
 is degenerate to $f_V$ on the $\{V,H\}$ boundary.
Obviously, the number of such   structures is infinite as $K\to \infty$.
Note that (\ref{ord}) is a special case of (\ref{o2}).
This degeneracy has also 
been confirmed in our numerical analysis of (\ref{free1}).

Generally if $y_k=\pm 1$ for all layers, the free energy (\ref{free1})
can be explicitly evaluated for any ordering.  
Let $p_{\sigma, \sigma '}$, $\sigma =\pm $,
 denote the fraction of layers with $y_k=-1$ such that
$\{y_{k-1}, y_k, y_{k+1}\} = \{ \sigma, -, \sigma '\}$,
where for brevity we denote $\pm 1$ by $\pm$.
Consider, for example, the domain   $ v<w$. 
A straightforward evaluation of (\ref{free1}) leads to the expression
\begin{eqnarray}
f&=& \a_+ \ln u  + \a_- \ln w, \hskip 3.7cm    ve^{-4h} < ve^{4h}<w \nonumber \\
   &=& \a_+ \ln u + \bigl(p _{++}+p _{-+}+p _{--}\bigr)\ln v \nonumber\\
&& \hskip 1.5cm
+ p _{+-}\ln w + 4h p _{-+} \hskip 0.8cm
    ve^{-4h} < w <  ve^{4h}.  \label{fh5}
\end{eqnarray}
The  first line can 
 be discarded since it is always smaller than the larger
of $\ln u$ and $\ln w$.
A degeneracy of states now occurs if the second line coincides with the free energy
of any phase.
In the case of the ordering (\ref{o2}), for example, it is translated to
$$
 \cdots -(+)----(+)------(+)--\cdots .
$$
Now  since each $+$ layer is associated with 
precisely one $+-$ and one $-+$ neighbors, we have
 $\a_+ = p_{+-} = p_{-+}$.  The 
 substitution of $\a_+ = p_{+-} = p_{-+}$ and
  (\ref{uvwh}) into (\ref{fh5})
now leads to $f=f_V$, in agreement with the  energy consideration.
 The degenerate states on the $\{H,U\}$ and $\{H,W\}$ boundaries can be
obtained by cyclic permutations of $u,v,w$.

The entropy of the ordered state (\ref{o2}) can be computed. We note that the 
main feature of (\ref{o2}) is that layers $v$ are followed by either $u$ or $v$ layers,
and $u$ layers are followed by only $w$ layers, and $w$ layers by $v$.
   Then   the degeneracy
$S$ of the sequence (\ref{o2}) is given by the trace of a transfer matrix as
\begin{equation}
S = {\rm Tr} \>(T^{K})=\la_1^K +\la_2^K  + \la_3^K \> \sim \la_1^K,
\hskip 1cm K \>\>{\rm large}
\end{equation}
where 
 \begin{equation}
T = \pmatrix{0&0&1\cr 1&0&1\cr 0&1&0\cr},
\end{equation}
and $\la_j$'s are the eigenvalues of  $T$ with $\la_1 > |\la_j|,\> j=2,3$.
We find
\begin{eqnarray}
\la_1 &=& {1\over 3}\biggl[
 1+  \biggl({29\over 2} + {{3 \sqrt{93}} \over 2}\biggr)^{1/3} +
\biggl({29\over 2} -{{3 \sqrt{93}} \over 2}\biggr)^{1/3} \biggr] \nonumber \\
&\sim& 1.46557... \nonumber 
\end{eqnarray}

\section{Summary}

We have considered a three-dimensional layered dimer system 
with interlayer interactions
and its equivalent layered domain wall
model, and analyzed its exact solution.
 It is found that the phase diagram, shown in Fig. 6,
 depends crucially on  the strength of
the interlayer interactions.  There exist ordered
$U,V,W,H$ phases corresponding to, respectively, large dimer weights
$u,v,w$ and large interlayer interactions $h$.
In addition, there also exist
a disorder phase $Y$ and partially
ordered phases $I_u, I_v, I_w$.  The phase boundaries 
are determined by equating the free energies of adjacent  regimes.  Particularly, 
the boundary between
the $U,V,W$ phases and the $Y$ phase assume the simple form  (\ref{yuvwboundary}),
between the $H$ and $Y$ phases  the form (\ref{hyboundary}),
and between the $H$ and $I_u$ phases the form (\ref{iuhboundary}).
All transitions are found to be of first order, 
except the transitions between the $U,V,W$ phases and the $Y$ phase,
and the transitions between the $I_u, I_v, I_w$ and $H$
phases, which are of second order with a square-root singularity 
in the specific heat.

\section{Acknowledgements}

Work by FYW and HYH has been  supported in part by National Science Foundation
Grants DMR-9313648 and DMR-9614170, work by VP has been supported in part by
INTAS Grants 93-1324 and 93-0633, and VP and 
DK by the Korea Science and Engineering
Foundation through the SRC program.

\begin{figure}
\caption{A three-dimensional lattice model consisting of layers
of honeycomb  dimer lattices.}
\label{5vlayer}
\end{figure}

\begin{figure}
\caption{The mapping of a honeycomb lattice onto a square lattice.}
\label{5vlbrick}
\end{figure}

\begin{figure}
\caption{Vertex configurations and weights of the 5-vertex model.}
\label{5vmodel}
\end{figure}

\begin{figure}
\caption{The three possible orientations of a diamond.
Strips are associated with diamonds oriented in two particular directions.}
\label{domain}
\end{figure}

\begin{figure}
\caption{The six strip configurations and the associated weights
of a triangle.}
\label{domain2}
\end{figure}

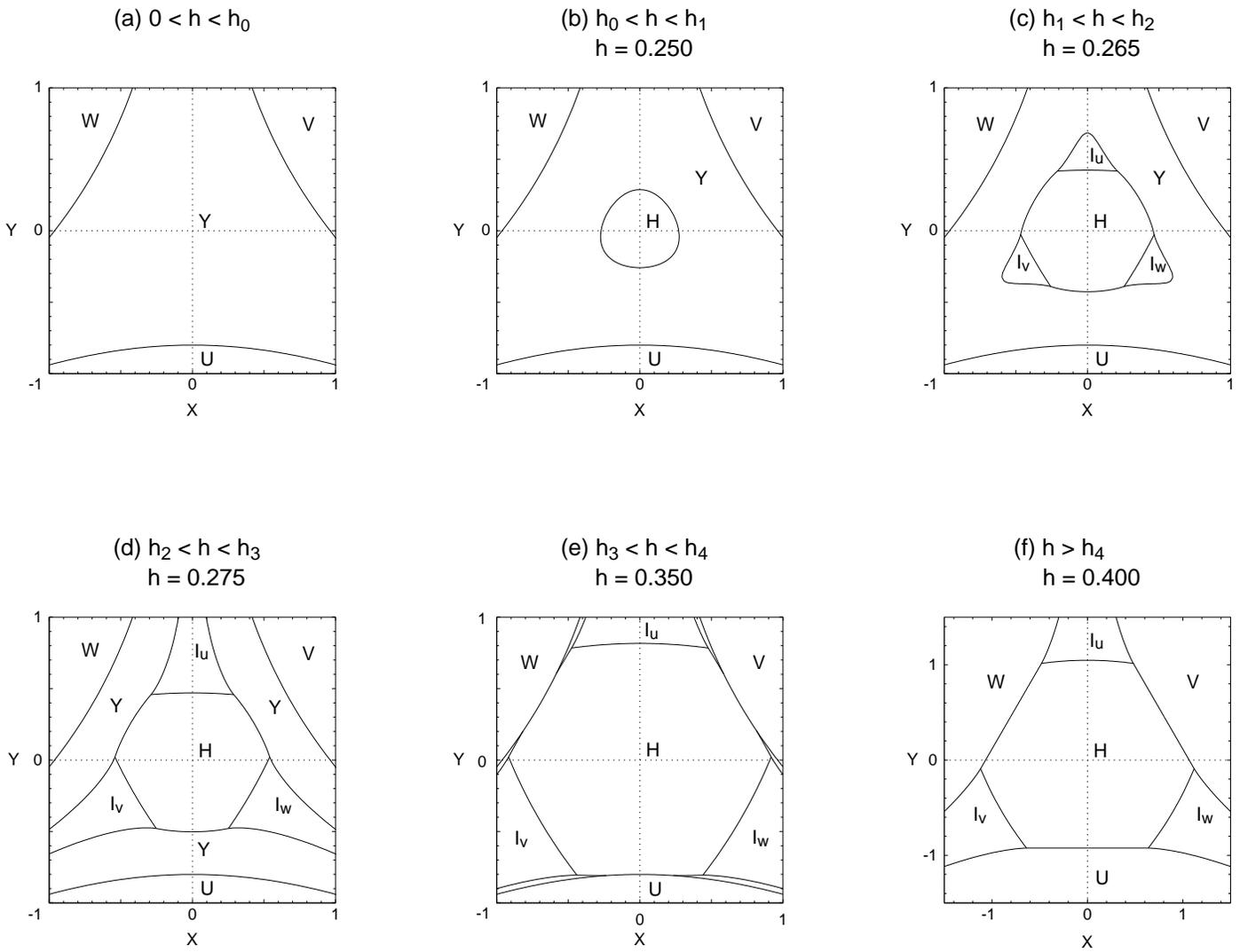
\begin{figure}
\caption{Phase diagram of the 3$D$ system.}
\label{5vlpha}
\end{figure}

\begin{table}
\caption{Interaction energy between two dimers  
	 incident at the same site of adjacent layers.
The interaction is symmetric in $u, v, w$.}
\smallskip

\begin{tabular}{cccc}
\multicolumn{1}{c}{layer $k\to k+1$}
&\multicolumn{1}{c}{$u$}
&\multicolumn{1}{c}{$v $}
&\multicolumn{1}{c}{$w$}
\\
\tableline
$u$ & 0 & $2h/3$  & $-2h/3$ \\
$v$ & $-2h/3$&  0 & $2h /3$ \\
$w$ & $2h/3$ & $-2h/3$ & 0 \\
\end{tabular}
\end{table}
 
\begin{table}
\caption{Interaction energy between two vertex
configurations of adjacent layers. $\omega_i, i=2,\cdots,6$
denotes the vertex of type $i$ in Fig. 3.}
\smallskip

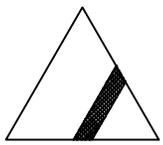
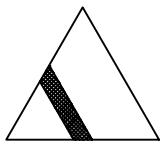
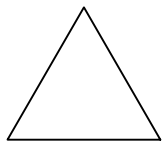
\begin{tabular}{cccccc}  
layer $k-k+1$ & $\omega_2$ & $\omega_3$ & $\omega_4$ & 
      $\omega_5$ & $\omega_6$ \\ \hline
$\omega_2$ & $0$ & $-4h/3$ & $4h/3$ & $0$ & $0$ \\ 
$\omega_3$ & $4h/3$ & $0$ & $-4h/3$ & $4h/3$ 
           & $-8h/3$ \\ 
$\omega_4$ & $-4h/3$ & $4h/3$ & $0$ & $-4h/3$ 
           & $8h/3$ \\ 
$\omega_5$ & $0$ & $8h/3$ & $-8h/3$ & $0$ & $0$ \\ 
$\omega_6$ & $0$ & $-4h/3$ & $4h/3$ & $0$ & $0$  \\ 
\end{tabular}
\end{table}

\begin{table}
\caption{Interaction energy between two strip triangles
 of adjacent layers. The triangle configurations are as numbered in Fig. 5.}
\smallskip

\begin{tabular}{ccccccc}
\multicolumn{1}{c}{layer $k\to k+1$}
&\multicolumn{1}{c}{1}
&\multicolumn{1}{c}{2}
&\multicolumn{1}{c}{3}
&\multicolumn{1}{c}{4}
&\multicolumn{1}{c}{5}
&\multicolumn{1}{c}{6}
\\
\tableline
1 & 0 & 0  & 0 &0 &$ -2h$ & 0\\
2 & 0 &  0 & 0& $2h$ & 0 & 0 \\
3 &0&0&0&0&0&0 \\
4 & $0$ & $-2h$ &0&0 & 0 &0 \\
5& $2h$ & 0 &0& 0 & 0&0 \\
6 &0&0&0&0&0&0 \\
\end{tabular}
\end{table}


\begin{references}
\bibitem{remark1} Except  on the  boundaries between the $H$ phase
and the $U,V,W$ phases other  layer orderings 
also set in.  The occurrence of this degeneracy is discussed in Sec. VII.
\bibitem{remark} The occurrence of $h_3=(\ln 2)/2$ was not discussed in
\cite{5vl1}.
\end{references}
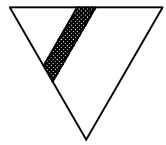
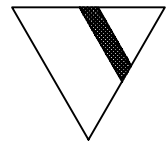
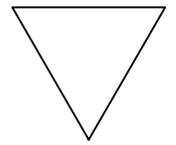
\end{document}